\documentclass[aps,amsmath,amssymb,superscriptaddress,nofootinbib]{revtex4}

\usepackage{graphicx}
\usepackage{hyperref}
\usepackage{epsf}
\usepackage{bm}
 \usepackage{slashed}
\usepackage{hyperref}
\usepackage{enumitem}
\usepackage{lipsum}
\usepackage{bbold}
\usepackage{sidecap}
\usepackage{subfigure}
\usepackage{multirow}

\newcommand{\bea}{\begin{eqnarray}}
\newcommand{\eea}{\end{eqnarray}}
\newcommand{\la}{\label}
\newcommand{\be}{\begin{equation}}
\newcommand{\ee}{\end{equation}}

\def\12{\frac{1}{2}}

\newcommand{\tr}{\,\mbox{tr}\,}

\def\XXint#1#2#3{{\setbox0=\hbox{$#1{#2#3}{\int}$}
     \vcenter{\hbox{$#2#3$}}\kern-.5\wd0}}

\numberwithin{equation}{section}

\begin{document}

\title{A Stringy (Holographic) Pomeron with Extrinsic Curvature}

\author{Yachao Qian}
\address{Department of Physics and Astronomy,
Stony Brook University,  Stony Brook, NY 11794-3800.}

\author{Ismail Zahed}
\address{Department of Physics and Astronomy,
Stony Brook University,  Stony Brook, NY 11794-3800.}

\begin{abstract}
We model the soft pomeron in QCD using a scalar Polyakov string with extrinsic curvature
in the bottom-up approach of holographic QCD. The overall dipole-dipole  scattering amplitude in 
the soft pomeron kinematics is shown to be sensitive to the extrinsic curvature of the string
for finite momentum transfer. The characteristics of the diffractive peak in the differential elastic $pp$ 
scattering are affected by a small extrinsic curvature of the string.
\end{abstract}
\date{\today}

\maketitle

\section{\label{sec:introduction}introduction}

The high energy proton on proton (anti-proton)  cross sections are dominated by Pomeron exchange,
an effective object corresponding to the highest Regge trajectory. The slowly rising cross sections
are described by the soft Pomeron with intercept $\alpha_P(0)-1\approx 0.08$ and vacuum quantum
numbers.  Reggeon exchanges have smaller intercepts and are therefore subleading. Reggeon theory 
for hadron-hadron scattering with large rapidity intervals provide an effective explanation for the transverse 
growth of the cross sections~\cite{Gribov:1984tu}.

The transverse growth of the proton with rapidity $\chi$ follows from the BFKL  ladders~\cite{Kuraev:1976ge,Lipatov:1976zz,Sterman:1999yc,Fadin:1975cb,Balitsky:1978ic} at weak coupling in QCD. Collinear gluon bremsstrahlung is large even when the coupling is weak
and requires re-summation. The ensuing BFKL hard Pomeron carries a large intercept and zero
slope. The intercept is slightly improved  by  higher order perturbative corrections to the BFKL ladder.

The soft Pomeron kinematics suggests an altogether  non-perturbative approach. Through duality arguments, 
Veneziano suggested long ago that the soft Pomeron is a closed string exchange~\cite{Veneziano:1968yb}. In QCD
the closed string world-sheet can be thought as the surface spanned by planar gluon diagrams or
fish-nets~\cite{Greensite:1984sb}.  The quantum theory of planar diagrams in supersymmetric gauge
theories is tractable in the double limit of a large number of colors $N_c$ and ${}^{\prime}$~t Hooft coupling
$\lambda=g^2N_c$ using the AdS/CFT holographic approach~\cite{Maldacena:1998im}.

In the past decade there have been several attempts at describing the soft pomeron using
holographic QCD~\cite{Rho:1999jm,Janik:2000aj,Janik:2000pp,Polchinski:2001tt,Polchinski:2002jw,Brower:2006ea,Brower:2007xg,Brower:2010wf,Brower:2011dx,Hatta:2007cs,Hatta:2007he,Albacete:2008ze,Albacete:2008vv,Basar:2012jb,Stoffers:2012zw,Stoffers:2012ai,Stoffers:2013tla,Zahed:2012sg,Shuryak:2013sra}.  In this letter we follow the 
work in~\cite{Stoffers:2012zw,Stoffers:2012ai,Stoffers:2013tla} and describe the soft pomeron as an effective string with extrinsic curvature in 5-dimensions. This is inherently a bottom-up approach with the holographic or 5th direction
playing the role of the scale dimension for the closed string. The geometry is that of AdS$_5$ with a wall. In the UV AdS$_5$ enforces conformality which is a property of QCD-BFKL-kernels, while in the IR the wall enforces confinement a generic
feature of QCD.

In section 2 we review the set up for dipole-dipole scattering through a closed string exchange.
In section 3, we introduce the QCD effective action with a finite string tension and extrinsic curvature and use it to derive the
closed string exchange propagator in flat $5=2+D_\perp$ dimensions. In section 4, we detail how the extrinsic curvature modifies 
the correlation of twisted Wilson loops, and show how it affects the position of the diffractive peak in the differential elastic
$pp$ scattering cross section. Our conclusions are in section 5. In the Appendix, we show how the extrinsic curvature affects the stringy interaction between two static dipoles.

\section{\label{scattering} Dipole-Dipole Scattering}

\begin{figure}[!htb]
 \includegraphics[height=50mm]{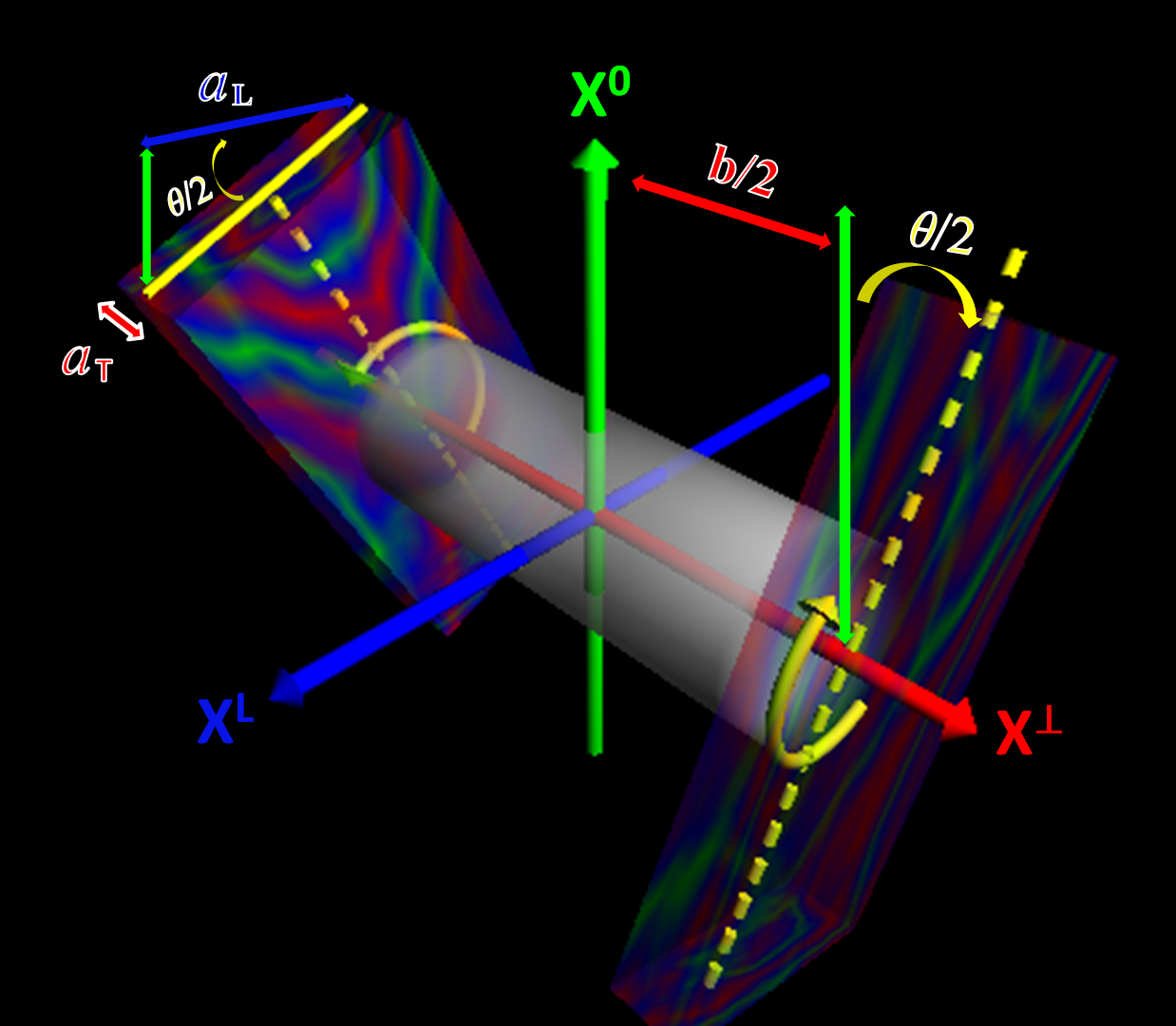}
  \caption{Dipole-Dipole Scattering.}\label{kinematics}
\end{figure}

In this section we briefly review the set-up for dipole-dipole scattering using an effective string theory.
For that we follow~\cite{Basar:2012jb} and consider the elastic scattering of two dipoles

\be
D_1 (p_1) + D_2 (p_2) \rightarrow D_1 (k_1) + D_2 (k_2)
\ee
as depicted in Fig.~\ref{kinematics}. $a_T$ and $a_L$ are the dipoles transverse and longitudinal lengths respectively
set near the UV boundary of AdS$_5$, $b$ is impact parameter and the angle $\theta$ is the Euclidean analogue of the rapidity interval

\be
\cosh \chi = \frac{s}{2 m^2} -1 \rightarrow\cos \theta
\ee
with $s = (p_1 + p_2)^2$.
Following the same argument as in~\cite{Basar:2012jb}, the scattering amplitude $\mathcal{T}$ in Euclidean space is given by

\be
\frac{1}{- 2 i s}  \mathcal{T} (\theta, q) \approx \int d^2 \bold{b}~  e^{i \bold{q}_\perp \cdot \bold{b} } \langle\bold{W}(-\frac{\theta}{2} , - \frac{b}{2}) \bold{W}(\frac{\theta}{2} ,  \frac{b}{2}) -1\rangle
\ee
with
\be
\bold{W} (\theta, \bold{b}) = \frac{1}{N_c} \tr [{\rm \bold{P}}_c \exp (i g \int_{\mathcal{C}_\theta}  d \tau ~ \bold{A} (\bold{x}) \cdot \bold{v})      ]
\ee
is the normalized Wilson loop for a dipole with $\langle \bold{W}\rangle=1$. $\mathcal{C}_\theta$ is the closed rectangular loop in Fig.~\ref{kinematics}. For simplicity, we denote the Euclidean loop correlator as $\bold{W}\bold{W}$. Considering one closed string exchange between dipoles, we have
\be\la{wwcalculation}
\langle\bold{W} \bold{W}\rangle \equiv  \langle \bold{W}(-\frac{\theta}{2} , - \frac{b}{2}) \bold{W}(\frac{\theta}{2} ,  \frac{b}{2}) -1\rangle = g_s^2 \int \frac{d T}{2 T} \bold{K} (T)
\ee 
where
\be\la{partitionfunction}
\bold{K} (T) = \int_{T}  \mathfrak{D}[x] ~ e^{- S[x] + {\rm ghost}}
\ee
is the string partition function on the cylinder topology with modulus $T$. The overall factor of 
$g_s^2$ in (\ref{wwcalculation})  is due to the relative genus in comparison to the unconnected Wilson loops. 
This analysis of the soft pomeron is different from the (distorted) spin-2 graviton exchange in~\cite{Brower:2006ea,Brower:2007xg,Brower:2010wf,Brower:2011dx}
as the graviton is massive in walled AdS$_5$. Our approach is similar to the one followed in~\cite{Basar:2012jb} with the difference
that $2+D_\perp=5$ and not 10~\cite{Stoffers:2012zw,Stoffers:2012ai,Stoffers:2013tla}. It is essentially an effective approach along the bottom-up scenario of AdS$_5$ with metric

\be
ds^2=\frac {R^2}{z^2}\left( (dx^0)^2+(dx^1)^2+ (dx_\perp)^2+(dz)^2\right)
\label{X1}
\ee
and $0\leq z\leq z_0$.  $R$ is the size of the AdS space for $z_0=\infty$.
Although the dual field theory corresponding to this truncated version of AdS$_5$ 
metric is not QCD, it does capture some key aspects, i.e. conformality in the UV and confinement in the IR. A similar argument was made in~\cite{Brodsky:2013npa}  in calculating the light-front wave-functions from the AdS/CFT holographic correspondence.

\section{\label{kappa} ${\bf K}$ with Extrinsic Curvature }

At large impact parameters ${\bf b}$ and for fixed dipole sizes $a$ on the boundary, the exchanged string in Fig.~\ref{kinematics}
is long and lies mostly along the wall at $z\approx z_0$ whereby the metric is nearly flat
~\cite{Janik:2000aj,Janik:2000pp,Basar:2012jb,Stoffers:2012zw,Stoffers:2012ai,Stoffers:2013tla,Zahed:2012sg,Shuryak:2013sra}.
In~\cite{Basar:2012jb,Stoffers:2012zw,Stoffers:2012ai,Stoffers:2013tla}, the authors used  the scalar Polyakov string action and showed that such a single closed string exchange yields a Regge behavior of the elastic amplitude.  We now revisit this analysis by considering the corrections due to the extrinsic curvature of the effective string action as advocated also by Polyakov~\cite{Polyakov:1986cs}.

\subsection{Effective String Action}

There are many indications from lattice simulations that flux tubes in Yang-Mills theory can be described by
an effective theory of strings of which the Nambu-Goto (NG) action is a good approximation in leading order~\cite{Kuti:2005xg}. Polyakov has suggested that the NG action must include an effective contribution that accounts for the extrinsic curvature of the world-sheet at
next order. The extrinsic curvature favors smooth string configurations and penalizes strings with high curvature. Specifically, the
scalar action in Polyakov form with extrinsic curvature is~\cite{Polyakov:1986cs,Hidaka:2009xh}

\be\la{action}
S[x] =  \frac{\sigma_T}{2} \int_0^T d \tau \int_0^1 d \sigma ~ (\dot{x}^\mu \dot{x}_\mu + {x'}^\mu {x'}_\mu ) 
+  \frac{1}{2 \kappa b^2} \int_0^T d \tau \int_0^1 d \sigma ~ ( \ddot{x}^\mu \ddot{x}_\mu + 2 \dot{x'}^\mu  \dot{x'}_\mu + {x''}^\mu {x''}_\mu ) 
\ee
We have set the gauge on the world-sheet to be $h_b^a=\delta_b^a$
and used the nearly flat metric $g^{\mu\nu}(z\approx z_0)=\delta^{\mu\nu}R^2/z_0^2$ at the bottom of AdS$_5$ (long strings).
Here $\dot{x}=\partial_\tau x$ and $x^\prime=\partial_\sigma x$. The string tension is $\sigma_T=1/(2\pi \alpha^\prime)$
with $\alpha^\prime=z_0^2/\sqrt{\lambda}$, and the effective and dimensionless extrinsic curvature is $\kappa=ez_0^2/R^2$.

\subsection{Boundary Conditions}

For small dipole size and large impact parameter, the boundaries of the funnel of the exchanged string will be pinched and can be approximated as two straight lines

\bea\la{twistboundary}
\cos (\frac{\theta}{2}) x^1 + \sin (\frac{\theta}{2}) x^0  ~~~ |_{\sigma=0} &=& 0 \nonumber\\
\cos (\frac{\theta}{2}) x^1 - \sin (\frac{\theta}{2}) x^0  ~~~ |_{\sigma=1} &=& 0 
\eea
and $x^\mu$ is periodic along the $\tau$ direction
\be
x^\mu (\tau) = x^\mu (\tau + T)
\ee
The twisted boundary condition (Eq.~\ref{twistboundary}) can be simplified as follows

\bea
\begin{pmatrix} x^0   \\  x^1 \end{pmatrix} =
 \begin{pmatrix}
 \cos \frac{\theta_\sigma}{2}  & - \sin \frac{\theta_\sigma}{2}  \\ 
\sin \frac{\theta_\sigma}{2}  & \cos \frac{\theta_\sigma}{2} 
 \end{pmatrix}
 \begin{pmatrix} 
y^0 
 \\  
y^1 
\end{pmatrix} 
\eea
with $\theta_\sigma = \theta (2 \sigma - 1)$. As a result (\ref{twistboundary}) are now ordinary 
Dirichlet boundary conditions
 
\be\la{dirichlet}
y^1 ~~ |_{\sigma=0 , 1}=0
\ee
By taking $\partial_\tau$ on both sides of  Eq.~\ref{dirichlet} and recalling that
the world-sheet energy-momentum tensor is null,  i.e. 
$T^{\alpha \beta} = \delta S / \delta g_{\alpha \beta} = 0$, we have

\bea
\partial_\tau y^1 ~~ |_{\sigma=0 , 1} &=& 0 \nonumber\\
\partial_\sigma y^0 ~~ |_{\sigma=0 , 1} &=& 0
\eea

\subsection{Closed String Propagator {\bf K}}

The natural mode decomposition for the string cordonates

\bea\la{fourierdecompose}
y^0 (\tau, \sigma) &=& \sum_{m=-\infty}^\infty \sum_{n=0}^{\infty}  y_{m,n}^0 \exp({i 2 \pi m \frac{\tau }{T}}) \cos (\pi n \sigma) \nonumber\\
y^1 (\tau, \sigma) &=& \sum_{m=-\infty}^\infty \sum_{n=1}^{\infty}  y_{m,n}^1 \exp({i 2 \pi m \frac{\tau }{T}}) \sin (\pi n \sigma) \nonumber\\
x^\perp (\tau, \sigma) &=& y^\perp (\tau, \sigma) = (\sigma - \frac{1}{2}) b^\perp  +  \sum_{m=-\infty}^\infty \sum_{n=1}^{\infty}  y_{m,n}^\perp \exp({i 2 \pi m \frac{\tau }{T}}) \sin (\pi n \sigma) 
\eea
with $b^\perp=(0,0,b,0,0)$ along one of the 2 spatial perpendicular directions. A rerun of the arguments presented in
~\cite{Basar:2012jb,Zahed:2012sg} yield  the closed string propagator ${\bf K}$ in~\ref{partitionfunction} in the form

\be\la{propa}
\bold{K} = \bold{K}_{0L} \times \bold{K}_{\O L} \times \bold{K}_\perp \times \bold{K}_{\rm ghost}
\ee
$\bold{K}_{0L}$ and $\bold{K}_{\O L}$ are the longitudinal zero and non-zero mode contributions respectively, $\bold{K}_\perp $ is the transverse contribution, and  $ \bold{K}_{\rm ghost}$ is the ghost contribution. Their explicit forms are 

\be\la{kol}
\bold{K}_{0L} = \{ \prod_{m=-\infty}^\infty  [ {  \frac{\sigma_T T}{2 \pi}     ( \theta^2 + \frac{4 \pi^2 m^2}{T^2}) + \frac{T}{2 \kappa b^2 \pi}   ( \theta^2 + \frac{4 \pi^2 m^2}{T^2})^2 }] \}^{-\frac{1}{2}}
\ee
\be\la{knotol}
\bold{K}_{\O L} =\{ \prod_{n=1}^{\infty}  \prod_{s=\pm}   \prod_{m=-\infty}^\infty {  ~ [  \frac{\sigma_T T }{4 \pi}     ( \frac{4 m^2 \pi^2 }{T^2} +    (n \pi+ s \theta )^2)    + \frac{T}{4 \kappa b^2\pi}  ( \frac{4 m^2 \pi^2 }{T^2} +    (n \pi+ s \theta )^2) ^2 ]  } \}^{-\frac{1}{2}}
\ee
\be\la{kperp}
\bold{K}_\perp = \exp[- \frac{\sigma_T}{2} T b^2 ]  \{ \prod_{n=1}^{\infty}   \prod_{m=-\infty}^\infty  [\frac{\sigma_T T}{4 \pi} (\frac{4\pi^2 m^2}{T^2} + n^2 \pi^2) + \frac{T}{4 \kappa b^2 \pi} (\frac{4\pi^2 m^2}{T^2} + n^2 \pi^2)^2 ]     \}^{-\frac{D_\perp}{2}} 
\ee
\be\la{kghost}
\bold{K}_{\rm ghost} =  \prod_{n=1}^{\infty}   \prod_{m=-\infty}^\infty {  ~ [  \frac{\sigma_T T }{4 \pi}     ( \frac{4 m^2 \pi^2 }{T^2} + n^2 \pi^2 )    + \frac{T}{4 \kappa b^2\pi}  ( \frac{4 m^2 \pi^2 }{T^2} +  n^2 \pi^2) ^2 ]  } 
\ee
which are seen to reduce to those in~\cite{Basar:2012jb,Zahed:2012sg} for $\kappa=\infty$. The
ghost contribution beyond the scalar Polyakov action and for finite extrinsic curvature is assumed so as
to cancel the $s=\pm 1$ spurious non-zero modes contribution from the longitudinal contribution for $\theta=0$.
This assumption while proved for $\kappa=\infty$ is now assumed for finite $\kappa$.

The string of diverging products can be regularized by standard zeta function regularization
\be
\sinh (\pi x) = \pi x \prod_{m=1}^\infty \left( 1 + \frac{x^2}{m^2}    \right)
\ee
in terms of which the string partition function (\ref{propa}) now reads

\bea\la{fullpropagator}
\bold{K} (T, \kappa) =&&  \frac{a^2}{\alpha'}e^{- \frac{\sigma_T}{2} T b^2}   \frac{ 1 }{2\sinh \left(\frac{\theta T}{2}\right) } \left[ \prod_{n=1}^\infty \prod_{s =\pm} \frac{\sinh \left(  \frac{n \pi T}{2}  \right)  }{\sinh \left( \frac{T (n \pi + s \theta)}{2} \right)   } \right]   \left[\prod_{n=1}^\infty 2   \sinh \left(  \frac{n \pi  T}{2}    \right)     \right]^{-D_\perp} \nonumber\\
&\times& \frac{1}{2  \sinh \left(  \frac{T}{2} \sqrt{\theta^2 + \sigma_T \kappa b^2} \right)}  \prod_{n=1}^\infty \prod_{s =\pm} \frac{1  }{  2 \sinh \left[ \frac{T (n \pi + s \theta)}{2} \sqrt{1 +   \frac{\sigma_T \kappa b^2}{(n \pi + s \theta)^2}} \right] } \nonumber\\
&\times& 2^{D_\perp}  \left[2 \prod_{n=1}^\infty   \sinh \left(   \frac{n \pi T}{2}  \sqrt{ 1 +   \frac{\sigma_T \kappa b^2}{ n^2 \pi^2}}  \right)    \right]^{-D_\perp + 2}
\eea
with 
\be
a^2 \longrightarrow a_T^2 + \frac{a_L^2}{\sin^2 (\frac{\theta}{2})} \rightarrow   a_T^2
\ee
as the longitudinal dipole size $a_L$ is suppressed at large $\chi$ after analytical continuation.
Note that for large transverse impact parameter $b$ 

\bea
  \prod_{n=1}^\infty \prod_{s =\pm} \frac{  1 }{ 2 \sinh \left( \frac{T (n \pi + s \theta)}{2} \sqrt{1 +   \frac{\sigma_T \kappa b^2}{(n \pi + s \theta)^2}} \right) } 
  &\approx&  \exp \left[  -  \sum_{n=1}^\infty   \sum_{s = \pm} \frac{Tb   \sqrt{ \kappa  \sigma_T }}{2} \left(  1 + \frac{1}{2}   \frac{(n \pi + s \theta )^2}{ \sigma_T \kappa b^2}  + \cdots  \right)  \right] \nonumber\\
&\approx& \exp \left( - \zeta(0)   Tb   \sqrt{ \kappa  \sigma_T } \sqrt{1 + \frac{\theta^2}{\sigma_T \kappa b^2} } \right) \nonumber\\
&=& \exp \left(  \frac{T}{2}  \sqrt{\sigma_T \kappa b^2 + \theta^2} \right)
\eea
and 
\bea
 \prod_{n=1}^\infty 2 \sinh \left(  \frac{n \pi T}{2} \sqrt{1 +   \frac{\sigma_T \kappa b^2}{ n^2 \pi^2} } \right) &\approx&  \exp \left[ \sum_{n=1}^\infty  \frac{T b \sqrt{\kappa \sigma_T}}{2} \left( 1 + \frac{1}{2} \frac{n^2 \pi^2}{\sigma_T \kappa b^2}  + \cdots  \right)    \right] \nonumber\\
&=& \exp \left(  - \frac{T b \sqrt{\kappa \sigma_T}}{4}  \right)
\eea
where we used $\zeta(-2 n) = 0$ ($n = 1,2,3 , \cdots$). 
Thus (\ref{fullpropagator}) simplifies to

\be\la{propagator}
\bold{K} (T, \kappa) \approx  \bold{K}_{\rm F} (T)  \,\,    \exp \left[ \frac{(D_\perp -2)}{4} T b \sqrt{\kappa \sigma_T } \right]
\ee 
with $ \bold{K}_{\rm F} (T)$ the closed string propagator without the extrinsic curvature $\kappa$,

\be\la{freepropagator}
 \bold{K}_{\rm F} (T) = \frac{a^2}{\alpha'}   \frac{e^{- \frac{\sigma_T}{2} T b^2} }{ 2 \sinh \left(\frac{\theta T}{2}\right)  } 
  \prod_{n=1}^\infty \prod_{s =\pm} \frac{\sinh \left(  \frac{n \pi T}{2}  \right)    }{\sinh \left[ \frac{T (n \pi + s \theta)}{2} \right] }  
   \left[\prod_{n=1}^\infty 2  \sinh \left(  \frac{n \pi  T}{2}    \right)    \right]^{-D_\perp}  
\ee
The resulting (\ref{propagator}) is rather similar to the one derived in one-loop 
in~\cite{Hidaka:2009xh} for a large and static Wilson loop.
We now detail its impact on the scattering of two twisted dipoles with the soft Pomeron kinematics.

\section{\label{amplitude} Scattering Amplitude with Extrinsic Curvature}

\subsection{Dipole-Dipole Scattering}

The result (\ref{propagator}) may now be used to estimate the dipole-dipole scattering amplitude of section
\ref{scattering}. Indeed, inserting (\ref{propagator}) into (\ref{wwcalculation}),  and then
analytically continuing $\theta \longrightarrow - i \chi$, yield for the twisted Wilson-loop correlator

\be\la{ww1}
\bold{WW} =   \frac{ g_s^2 a^2}{4 \alpha'} \sum_{k = 1}^\infty \frac{(-1)^k}{ k }  e^{- k\frac{\pi \sigma_T \bold{b}^2}{\chi} }  \eta^{- D_\perp} (\frac{i k \pi }{\chi}) \,\,    \exp \left[ \frac{(D_\perp -2) k \pi }{2\chi}  b \sqrt{\kappa \sigma_T } \right]
\ee
where $\eta (\tau)$ is Dedekind eta function and $\eta (i x) =  \eta (i/x) /\sqrt{x}~ $\cite{Basar:2012jb,Zahed:2012sg} 
\be 
\eta^{- D_\perp} (\frac{i k \pi }{\chi})  =  (\frac{k \pi}{\chi})^{\frac{D_\perp}{2}} e^{\frac{ \chi D_\perp }{12 k }}  \prod_{n=1}^\infty (1 - \exp[- \frac{2  n \chi}{k } ])^{- D_\perp} = (\frac{k \pi}{\chi})^{\frac{D_\perp}{2}} e^{\frac{ \chi D_\perp }{12 k }}  \sum_{n=0}^\infty d(n) e^{-n \frac{2 \chi}{k}}
\ee 
 In momentum space, the scattering amplitude is
 
\bea\la{scatteringamplitude}
\frac{1}{- 2 i s}  \mathcal{T}_{DD} (\chi, q) &\approx& \int d^2 \bold{b}~  e^{i \bold{q}_\perp \cdot \bold{b} } \langle \bold{WW}\rangle \nonumber\\
&\approx&   \frac{\pi^2 g_s^2a^2}{2}  \sum_{n=0}^\infty  \sum_{k=1}^\infty ~ d(n)\frac{(-1)^k }{k} \left(\frac{k \pi}{\chi}\right)^{\frac{D_\perp - 2}{2}} \exp \left(\chi \frac{D_\perp}{12 k}- \chi \frac{2n}{k}  - \chi \frac{\bold{q}_\perp^2}{4 \sigma_T k \pi} + \sqrt{\kappa} \frac{D_\perp -2}{4} \frac{|\bold{q}_\perp|}{\sqrt{\sigma_T}} \right) \nonumber\\
\eea
where  we used $\chi \approx \ln s$ for large $s$. Recall that the sum over $k$ runs over the N-ality of the gauge group
which is up to $[N_c/2]=\infty$ in the AdS/CFT correspondence~\cite{Zahed:2012sg}. 
For QCD $N_c=3$ and $[3/2]=1$ which means only the
$k=1$ term contributes to the scattering of two dipoles in the fundamental representation of SU$(N_c)$. The effect of
the extrinsic curvature is a momentum dependent contribution to the exponent that is large but sub-leading at large
$\chi$.

\subsection{$pp$ Scattering}

For fixed impact parameter, $ \langle \bold{WW}\rangle$ is the elastic amplitude of a dipole of size $a$
onto a fixed dipole $a^\prime=a$, both of which are fixed in the UV or on the boundary. In general, the dipole size in a given 
hadron, say $p$ or $\bar p$ is scale dependent and identified with the holographic direction 
i.e. $a\rightarrow z=z_0e^{-u(z)}$ and $a^\prime\rightarrow z^\prime=z_0e^{-u^\prime(z^\prime)}$ with $0<u,u^\prime<\infty$~\cite{Stoffers:2012ai}. With this in mind the elastic scattering amplitude for $pp$ scattering reads in general

\be
\mathcal{T}_{pp}(\chi, \bold{b} ) = \int_0^\infty  d u \int_0^\infty d u'~ \psi_1^* (u) \psi_2^*(u')\mathcal{T}_{DD} \left(\chi,\bold{b} , u, u'\right) \psi_1 (u) \psi_2(u')
\ee 
In our case $|\psi_{1,2}(u,u^\prime)|^2\equiv {\cal N}_p\delta(u,u^\prime-u(a))$ for equal and fixed size dipoles $a$.
In the eikonal approximation the elastic differential cross section reads

\bea
\frac{d \sigma}{d t} =\frac{1}{16\pi s^2} \left|  \mathcal{T}_{pp} (\chi, q )  \right|^2 =\frac{1}{4 \pi}  \left|i \int d \bold{b}^2  \int du \int d u' \,\, e^{i \bold{q}_\perp \cdot \bold{b}} |\psi_1(u)|^2 |\psi_2(u')|^2  \left( 1 - e^{\bold{WW}}  \right)     \right|^2
\label{pp}
\eea
An optimal analysis of the available elastic differential $pp$ data  follows by setting: $D_\perp=3$, $N_c=3$, 
$\lambda=g^2N_c=9.4$, $\kappa_g=4\pi\,g_s/g^2=2.85$, $z_0=R=0.4$ fm,  ${\cal N}_p=1.5$ and $a=0.25$ fm
with a fixed rapidity interval $\chi=6$. This parameter set is overall consistent with the one used in~\cite{Stoffers:2012ai}
for the analysis of the DIS data. The results are displayed in Fig~\ref{crosssectionfigure} and compared to the elastic
$pp$ data for $\sqrt{s}=30.7, 44.7, 52.8$ GeV from~\cite{Amaldi:1971kt}. The dashed (blue) curve is for no extrinsic curvature $e=\kappa=0$ and the solid curve is for $e=\kappa=0.002$.  The slope parameter ${\bf B}(t)$ for the elastic differential cross section

\be
\bold{B} (t) = \frac d{dt}\left({\rm ln}\left(\frac {d\sigma}{dt}\right)\right)
\ee
is tabulated in~Table-\ref{slopetable}. While ${\bf B}(t)$ does not change with a small change in the extrinsic curvature
$e$, Fig~\ref{crosssectionfigure} shows that the depth and somehow the position of the diffractive peak are affected by
a small extrinsic curvature for a stringy description of the pomeron.  While  a more exhaustive analysis of the 
parameter space together with a better description of the dipole-dipole scattering amplitude at larger $|t|$ are needed, our
estimates show an interesting interplay between the characteristics of the diffractive peak and the extrinsic curvature of the
stringy pomeron.

\begin{table}
\label{slopetable}
 \begin{tabular}{|c|c|c|c|c|}
\hline
\multicolumn{1}{ |c  }{\multirow{2}{*}{ \ \  $\sqrt{s}$ [GeV]  \ \   } } & \multicolumn{1}{ |c |  }{\multirow{2}{*}{\ \ \ \ t [${\rm GeV}^2$] \ \ \ \  } }   & $\bold{B} (t)$    [${\rm GeV}^{-2}$]  &\multicolumn{2}{ | c| }{ $\bold{B} (t)$    [${\rm GeV}^{-2}$] } \\ \cline{4-5}
\multicolumn{1}{ |c  }{  } & \multicolumn{1}{ |c | }{  } & \ \ Experimental Data~\cite{Amaldi:1971kt} \ \  & \ \ \ \ \  $e = 0$ \ \ \ \ \  & \ \ $e = 0.002$ \ \ \\ \hline
30.7 & 0.015 - 0.055 & $13.0 \pm 0.7$ & 8.4 & 8.5 \\ \hline
44.7 & 0.03 - 0.15 & $12.9 \pm 0.4$ & 8.4 & 8.5 \\ \hline
52.8 & 0.04 - 0.16 & $13.0 \pm 0.3$ & 8.5 & 8.5 \\ \hline
\end{tabular}
\caption{Slope parameter ${\bf B}(t)$ for the elastic differential cross section.}
\end{table}

\begin{figure}[!htb]
\minipage{0.33333333333 \textwidth}
\includegraphics[height=70mm]{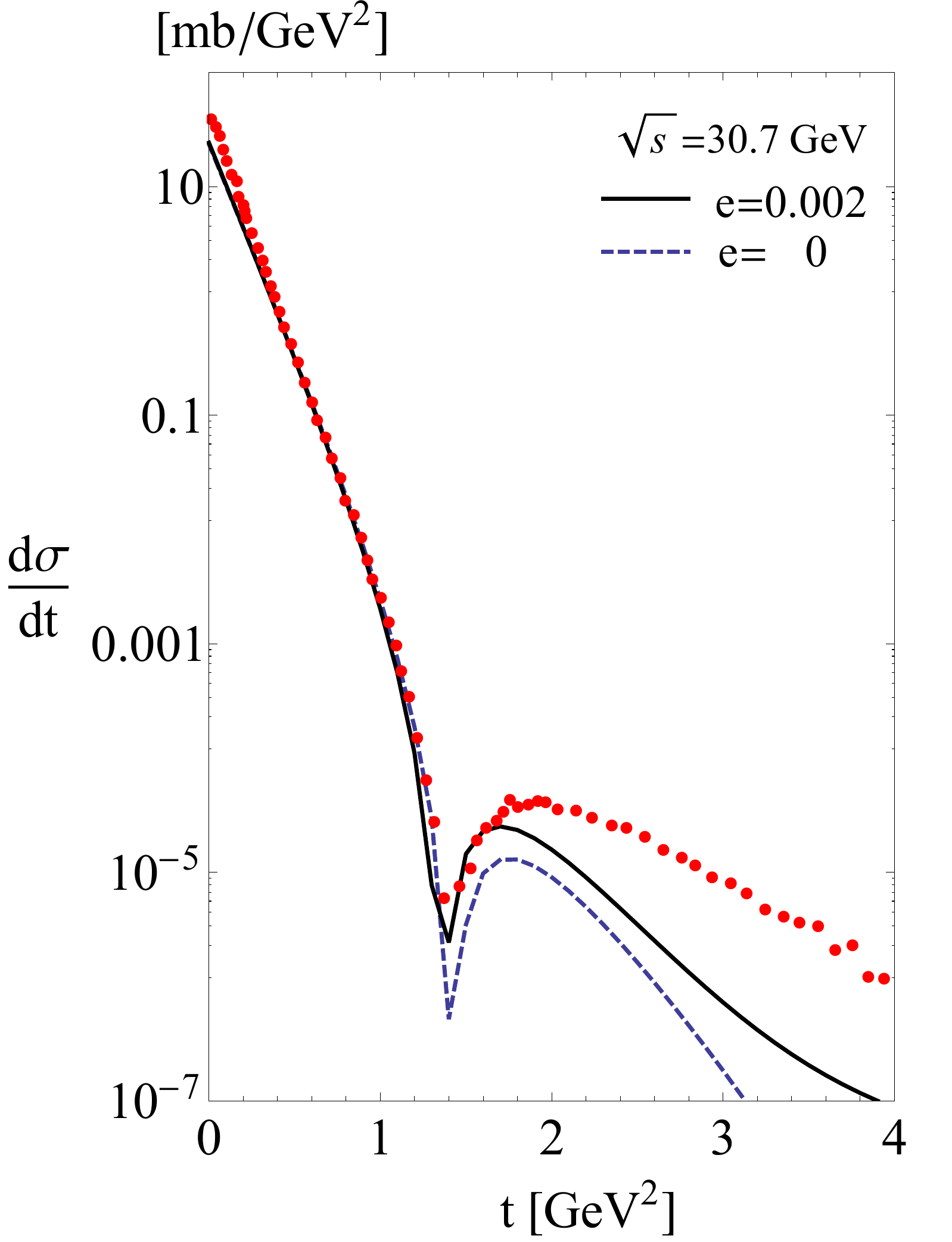}
\endminipage\hfill
\minipage{0.33333333333 \textwidth}
\includegraphics[height=70mm]{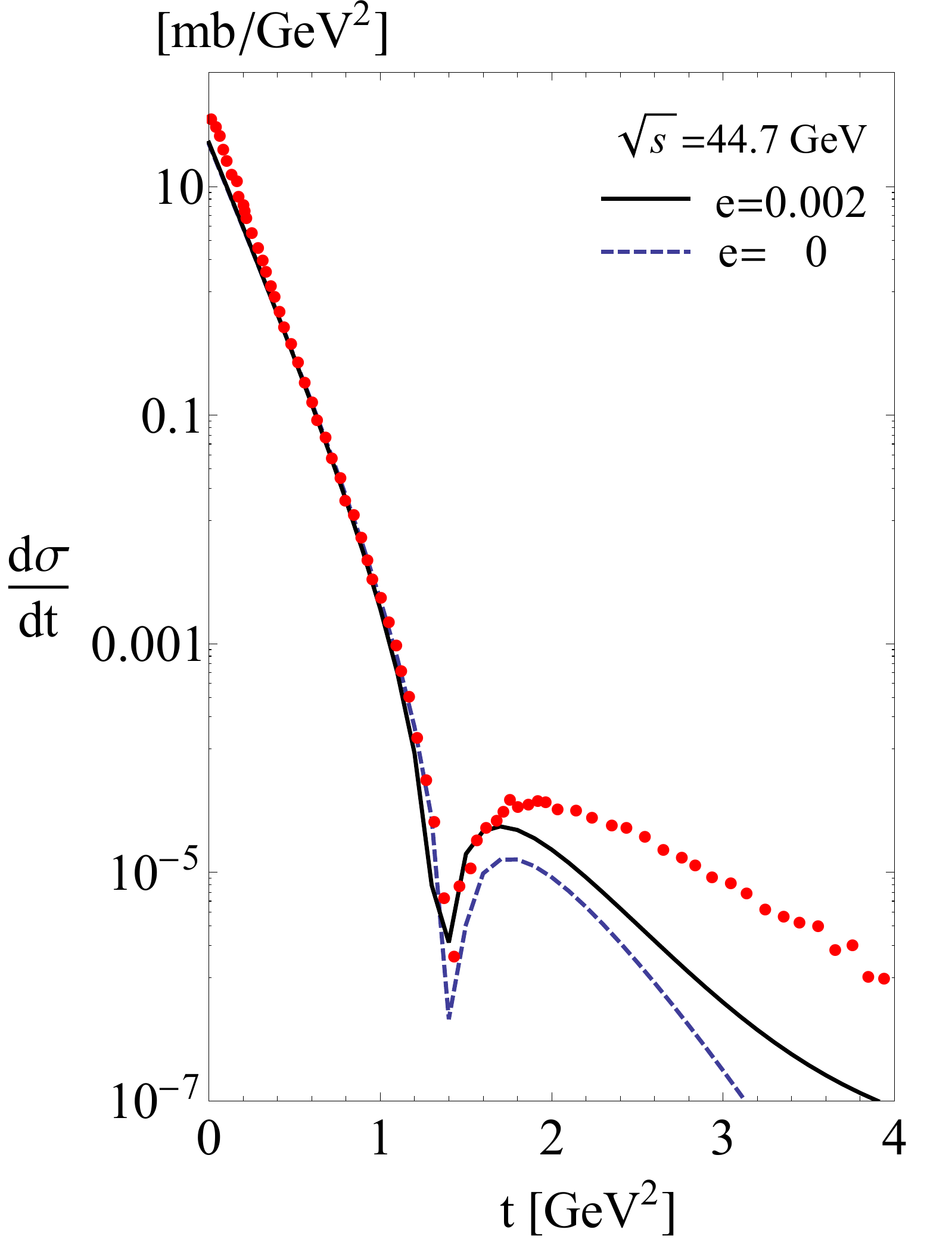}
\endminipage
\minipage{0.33333333333 \textwidth}
\includegraphics[height=70mm]{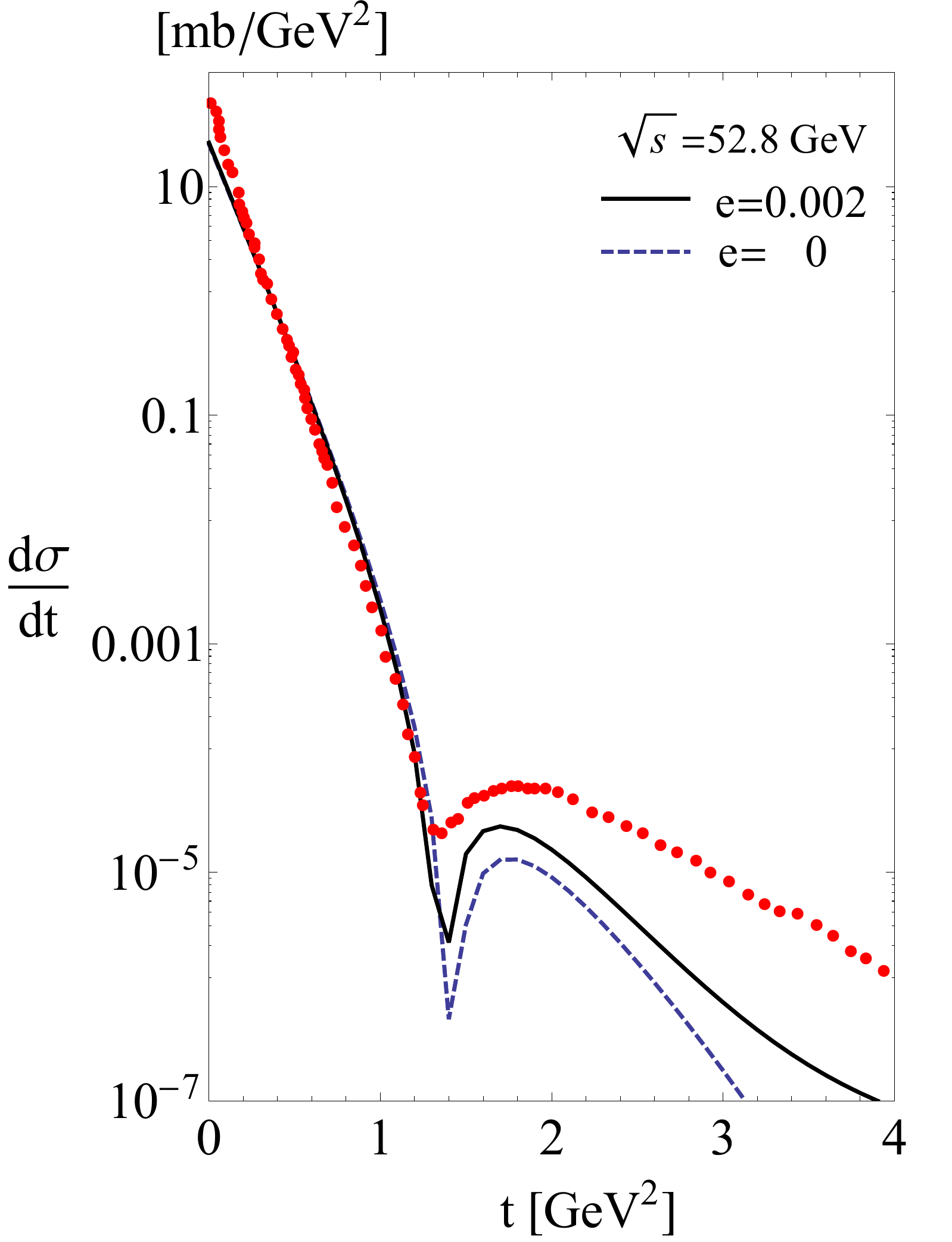}
\endminipage
  \caption{ Elastic differential $pp$ cross section: solid (black) curve stringy pomeron with 
  extrinsic curvature $\kappa=e=0.002$; dashed (blue) curve with $\kappa=e=0$;
  the data (red) is from~\cite{Amaldi:1979kd}.\la{crosssectionfigure}.}
\end{figure}

\section{\label{conclusion} Conclusion}

In holographic QCD the Pomeron exchange in dipole-dipole scattering with a large rapidity $\chi$
is described by the exchange of a non-critical string in hyperbolic $D=5$ dimensions. The extra
(curved) direction is identified with the string scale dimension. In leading order, the Pomeron intercept is set by 
the Luscher-like term or $D_\perp/12$~\cite{Basar:2012jb}, and its slope is set by the string tension 
at the confinement scale. The curvature of the extra dimension causes the Pomeron intercept to
shift from the Luscher term to order $1/\sqrt{\lambda}$~\cite{Stoffers:2012zw,Stoffers:2012ai,Stoffers:2013tla}.

Long color flux tubes in QCD are smooth. In leading order, the Nambu-Goto effective theory is corrected
by a term that depends on the extrinsic curvature to allow for smooth string configurations~~\cite{Polyakov:1986cs}.
The extrinsic curvature affects the zero point energy of large Wilson loops to one loop~\cite{Hidaka:2009xh} and
 is amenable to lattice simulations. We have shown that a similar contribution affects the 
scattering amplitude of two dipoles. While there are higher order (loop)
corrections to (\ref{scatteringamplitude}), the retained contributions are leading in the Pomeron kinematics.

In leading order, the extrinsic curvature induces an overall momentum 
dependent contribution to the scattering amplitude. Detailed comparison with accurate but differential 
proton on proton measurements at large $\sqrt{s}$ but fixed $t=-{\bf q}_\perp^2$ show sensitivity of the
diffractive peak to changes in the extrinsic curvature. $pp$ scattering may provide for an empirical
estimate of the extrinsic curvatures of smooth QCD strings, besides the current measurement estimates
for the slope (string tension) and intercept (Luscher contribution) of the Pomeron.

\section{\label{acknowledgements} Acknowledgements}

This work was supported by the U.S. Department of Energy under Contract No.
DE-FG-88ER40388.

\section{\label{zeropoint} Appendix: Static and Stringy Dipole-Dipole Interaction}

In this Appendix we detail the role of the extrinsic curvature on the correlator of two static but untwisted Wilson loops
with $\theta=0$, i.e. the interaction between two static dipoles. Instead of (\ref{fourierdecompose}) we now have the mode decomposition

\bea\la{untwistdecompose}
x^0 (\tau, \sigma) &=& \sum_{m=-\infty}^\infty \sum_{n=1}^{\infty}  x_{m,n}^0 \exp({i 2 \pi m \frac{\tau }{T}}) \cos (\pi n \sigma) + X + \frac{P}{\sigma_T} \tau   \nonumber\\
x^1 (\tau, \sigma) &=& \sum_{m=-\infty}^\infty \sum_{n=1}^{\infty}  x_{m,n}^1 \exp({i 2 \pi m \frac{\tau }{T}}) \sin (\pi n \sigma)  \nonumber\\
x^\perp (\tau, \sigma) &=&  (\sigma - \frac{1}{2}) b^\perp +  \sum_{m=-\infty}^\infty \sum_{n=1}^{\infty}  x_{m,n}^\perp \exp({i 2 \pi m \frac{\tau }{T}}) \sin (\pi n \sigma) 
\eea
with $P$ the number of windings in the temporal direction. The exchanged closed string is assumed
to be infinitely thin in this case in the absence of the boosting kinematics for the two scattering dipoles in the text.
This approximation is justified in the final result (\ref{FFXX}-\ref{FFXXX}) below. With this in mind, 
a repeat of the algebra in section \ref{kappa}, yields the string partition function 

\be\la{untwistpropagator}
\bold{K}  (T, \kappa) =  \bold{K}_{\rm F} (T ) \,\,  \exp\left(  \frac{D_\perp}{4} T b \sqrt{\kappa \sigma_T} \right)
\ee
with $ \bold{K}_{\rm F} (T)$ the string propagator propagator without the extrinsic curvature in \ref{action}
  
\be\la{untwistfreepropagator}
\bold{K}_{\rm F} (T ) = \frac{a^2}{\alpha'}\exp\left( - \frac{\sigma_T}{2} T b^2 - \frac{T P^2}{2 \sigma_T} \right)    \left[\prod_{n=1}^\infty 2   \sinh \left(  \frac{n \pi  T}{2}    \right)   \right]^{-D_\perp}
\ee
In comparing to the result in~\cite{Hidaka:2009xh}, we note the occurrence of the same zero point energy  (one loop)

\be
E_0^{\rm non}  = - \frac{D_\perp}{4} \sqrt{\sigma_T \kappa}  
\ee
This is to be compared with our result (\ref{propagator}) for the twisted dipoles, and shows the commonality between the
untwisted and large Wilson loop and the twisted and far Wilson loops.

Now, we also notice that in our case
\be
x^0 (\tau+T, \sigma) = x^0 (\tau , \sigma) + \frac{P}{\sigma_T} T
\ee
Thus $P = c W$ with $W=0, \pm 1, \pm 2, ...$ with $W$ the winding number and $c$ a constant to be interpreted below. 
 The propagators with different windings can be re-summed using the Poisson summation formula
 
\bea\la{windeq1}
   \sum_{W= -\infty}^\infty \bold{K}  (T, \kappa) &=&   \sum_{W= -\infty}^\infty \frac{a^2}{\alpha'}\exp\left( - \frac{\sigma_T}{2} T b^2 -  W^2   \frac{T  c^2 }{2 \sigma_T } +\frac{D_\perp}{4} T b \sqrt{\kappa \sigma_T} \right)    \left[\prod_{n=1}^\infty 2   \sinh \left(  \frac{n \pi  T}{2}    \right)   \right]^{-D_\perp} \nonumber\\
&=& \sqrt{\frac{2 \pi \sigma_T}{T c^2}} \frac{a^2}{\alpha'}  \sum_{k = -\infty}^\infty \exp\left[ - \frac{\sigma_T}{2} T b^2 \left( 1 - \frac{D_\perp \sqrt{\kappa}}{2 b \sqrt{\sigma_T}}   \right)-  k^2 \frac{2  \pi^2 \sigma_T}{ T c^2} \right]    \eta^{- D_\perp} \left( i \frac{T}{2}  \right)
\eea
where $\eta (\tau)$ is Dedekind eta function and $\eta (i x) =  \eta (i/x) /\sqrt{x}~ $\cite{Basar:2012jb,Zahed:2012sg}.  Inserting(\ref{windeq1}) into  (\ref{wwcalculation}) yields

\bea
\label{FFXX}
\bold{WW} &=&  \frac{g_s^2 a^2}{4 \alpha' }\frac{\sqrt{\pi \sigma_T}}{|c|} \sum_{k = -\infty}^\infty \sum_{n=0}^\infty d(n) \int_0^\infty d T \,\, \left( \frac{T}{2}  \right)^{\frac{D_\perp -3}{2}} \nonumber\\
&& \ \ \ \ \ \ \ \ \ \  \ \ \ \ \ \ \ \ \ \ \ \ \ \ \ \ \ \ \ \ \ \ \ \ \ \ \   \times  \exp \left[  - \frac{T}{2}  \sigma_T b^2  \left( 1 - \frac{D_\perp \sqrt{\kappa}}{2 b \sqrt{\sigma_T}}   \right) -  \frac{2}{T} k^2 \frac{   \pi^2 \sigma_T}{   c^2} \left(1+  \frac{2 n \pi c^2}{k^2 \pi^2 \sigma_T}   -  \frac{  D_\perp c^2}{12 k^2 \pi  \sigma_T }  \right) \right] \nonumber\\
&=&  \frac{g_s^2 a^2\sqrt{  \sigma_T}}{ \alpha' }\left( \frac{\pi}{|c|} \right)^\frac{D_\perp}{2} \sum_{k = -\infty}^\infty \sum_{n=0}^\infty d(n) \left( \frac{  1+  \frac{2 n \pi c^2}{k^2 \pi^2 \sigma_T}   -  \frac{  D_\perp c^2}{12 k^2 \pi  \sigma_T }  }{   1 - \frac{D_\perp \sqrt{\kappa}}{2 b \sqrt{\sigma_T}}   }  \right)^{\frac{D_\perp-1}{4}} \left( \frac{k}{b } \right)^{\frac{D_\perp-1}{2}} \nonumber\\
&&  \ \ \ \ \ \ \ \ \ \  \ \ \ \ \ \ \ \ \ \ \ \ \ \ \ \ \ \ \ \ \ \ \ \ \ \ \   \times    \bold{K}_{\frac{D_\perp-1}{2}} \left( \frac{2 b k \pi \sigma_T}{|c|} \sqrt{ \left( 1 - \frac{D_\perp \sqrt{\kappa}}{2 b \sqrt{\sigma_T}}   \right)  \left(1+  \frac{2 n \pi c^2}{k^2 \pi^2 \sigma_T}   -  \frac{  D_\perp c^2}{12 k^2 \pi  \sigma_T }  \right) }  \right)
\eea
which is the correlator between two static dipoles at large distances $b\gg D_\perp\sqrt{\kappa/4\sigma_T}$.
The summation over $k$ should be limited to 
$k=[N_c/2]=1$ for dipoles in the fundamental representation of SU(N$_c$)~\cite{Zahed:2012sg}.
$d(n)$ is the canonical string density of states with $d(0)=1$.
The static dipole-dipole potential following from the smooth  string exchange,  amounts to a tower
of scalar exchanges with masses ($k=1$)

\be
m_n(b,c)= \frac{2\pi \sigma_T}{|c|} \sqrt{ \left( 1 - \frac{D_\perp \sqrt{\kappa}}{2 b \sqrt{\sigma_T}}   \right)  \left(1+  \frac{2 n \pi c^2}{\pi^2 \sigma_T}   -  \frac{  D_\perp c^2}{12\pi  \sigma_T }  \right) }   
\ee
at large distances $b\gg D_\perp\sqrt{\kappa/4\sigma_T}$. Without the extrinsic curvature and setting $|c|\equiv 2\pi/\beta$, 
$m_n$ is the mass spectrum for closed strings of (arbitrary) size $\beta>\beta_H$, 
\be
\label{FFXXX}
m_n(\infty,2\pi/\beta)=\sigma_T\beta\sqrt{1-\frac{\beta_H^2}{\beta^2}+\frac{8\pi n}{\sigma_T\beta^2}}
\ee
with the Hagedorn  temperature $\beta_H=\sqrt{\pi D_\perp/3\sigma_T}$. Here $1/\beta=|c|/2\pi$ plays the role of an effective temperature associated with the exchange of a closed (periodic) string.


\bibliography{stringref}

\begin{thebibliography}{34}
\expandafter\ifx\csname natexlab\endcsname\relax\def\natexlab#1{#1}\fi
\expandafter\ifx\csname bibnamefont\endcsname\relax
  \def\bibnamefont#1{#1}\fi
\expandafter\ifx\csname bibfnamefont\endcsname\relax
  \def\bibfnamefont#1{#1}\fi
\expandafter\ifx\csname citenamefont\endcsname\relax
  \def\citenamefont#1{#1}\fi
\expandafter\ifx\csname url\endcsname\relax
  \def\url#1{\texttt{#1}}\fi
\expandafter\ifx\csname urlprefix\endcsname\relax\def\urlprefix{URL }\fi
\providecommand{\bibinfo}[2]{#2}
\providecommand{\eprint}[2][]{\url{#2}}

\bibitem[{\citenamefont{Gribov et~al.}(1983)\citenamefont{Gribov, Levin, and
  Ryskin}}]{Gribov:1984tu}
\bibinfo{author}{\bibfnamefont{L.}~\bibnamefont{Gribov}},
  \bibinfo{author}{\bibfnamefont{E.}~\bibnamefont{Levin}}, \bibnamefont{and}
  \bibinfo{author}{\bibfnamefont{M.}~\bibnamefont{Ryskin}},
  \bibinfo{journal}{Phys.Rept.} \textbf{\bibinfo{volume}{100}},
  \bibinfo{pages}{1} (\bibinfo{year}{1983}).

\bibitem[{\citenamefont{Kuraev et~al.}(1976)\citenamefont{Kuraev, Lipatov, and
  Fadin}}]{Kuraev:1976ge}
\bibinfo{author}{\bibfnamefont{E.~A.} \bibnamefont{Kuraev}},
  \bibinfo{author}{\bibfnamefont{L.~N.} \bibnamefont{Lipatov}},
  \bibnamefont{and} \bibinfo{author}{\bibfnamefont{V.~S.} \bibnamefont{Fadin}},
  \bibinfo{journal}{Sov.Phys.JETP} \textbf{\bibinfo{volume}{44}},
  \bibinfo{pages}{443} (\bibinfo{year}{1976}).

\bibitem[{\citenamefont{Lipatov}(1976)}]{Lipatov:1976zz}
\bibinfo{author}{\bibfnamefont{L.}~\bibnamefont{Lipatov}},
  \bibinfo{journal}{Sov.J.Nucl.Phys.} \textbf{\bibinfo{volume}{23}},
  \bibinfo{pages}{338} (\bibinfo{year}{1976}).

\bibitem[{\citenamefont{Sterman}(1999)}]{Sterman:1999yc}
\bibinfo{author}{\bibfnamefont{G.~F.} \bibnamefont{Sterman}}
  (\bibinfo{year}{1999}), \eprint{hep-ph/9905548}.

\bibitem[{\citenamefont{Fadin et~al.}(1975)\citenamefont{Fadin, Kuraev, and
  Lipatov}}]{Fadin:1975cb}
\bibinfo{author}{\bibfnamefont{V.~S.} \bibnamefont{Fadin}},
  \bibinfo{author}{\bibfnamefont{E.}~\bibnamefont{Kuraev}}, \bibnamefont{and}
  \bibinfo{author}{\bibfnamefont{L.}~\bibnamefont{Lipatov}},
  \bibinfo{journal}{Phys.Lett.} \textbf{\bibinfo{volume}{B60}},
  \bibinfo{pages}{50} (\bibinfo{year}{1975}).

\bibitem[{\citenamefont{Balitsky and Lipatov}(1978)}]{Balitsky:1978ic}
\bibinfo{author}{\bibfnamefont{I.}~\bibnamefont{Balitsky}} \bibnamefont{and}
  \bibinfo{author}{\bibfnamefont{L.}~\bibnamefont{Lipatov}},
  \bibinfo{journal}{Sov.J.Nucl.Phys.} \textbf{\bibinfo{volume}{28}},
  \bibinfo{pages}{822} (\bibinfo{year}{1978}).

\bibitem[{\citenamefont{Veneziano}(1968)}]{Veneziano:1968yb}
\bibinfo{author}{\bibfnamefont{G.}~\bibnamefont{Veneziano}},
  \bibinfo{journal}{Nuovo Cim.} \textbf{\bibinfo{volume}{A57}},
  \bibinfo{pages}{190} (\bibinfo{year}{1968}).

\bibitem[{\citenamefont{Greensite}(1985)}]{Greensite:1984sb}
\bibinfo{author}{\bibfnamefont{J.}~\bibnamefont{Greensite}},
  \bibinfo{journal}{Nucl.Phys.} \textbf{\bibinfo{volume}{B249}},
  \bibinfo{pages}{263} (\bibinfo{year}{1985}).

\bibitem[{\citenamefont{Maldacena}(1998)}]{Maldacena:1998im}
\bibinfo{author}{\bibfnamefont{J.~M.} \bibnamefont{Maldacena}},
  \bibinfo{journal}{Phys.Rev.Lett.} \textbf{\bibinfo{volume}{80}},
  \bibinfo{pages}{4859} (\bibinfo{year}{1998}), \eprint{hep-th/9803002}.

\bibitem[{\citenamefont{Rho et~al.}(1999)\citenamefont{Rho, Sin, and
  Zahed}}]{Rho:1999jm}
\bibinfo{author}{\bibfnamefont{M.}~\bibnamefont{Rho}},
  \bibinfo{author}{\bibfnamefont{S.-J.} \bibnamefont{Sin}}, \bibnamefont{and}
  \bibinfo{author}{\bibfnamefont{I.}~\bibnamefont{Zahed}},
  \bibinfo{journal}{Phys.Lett.} \textbf{\bibinfo{volume}{B466}},
  \bibinfo{pages}{199} (\bibinfo{year}{1999}), \eprint{hep-th/9907126}.

\bibitem[{\citenamefont{Janik and Peschanski}(2000)}]{Janik:2000aj}
\bibinfo{author}{\bibfnamefont{R.}~\bibnamefont{Janik}} \bibnamefont{and}
  \bibinfo{author}{\bibfnamefont{R.~B.} \bibnamefont{Peschanski}},
  \bibinfo{journal}{Nucl.Phys.} \textbf{\bibinfo{volume}{B586}},
  \bibinfo{pages}{163} (\bibinfo{year}{2000}), \eprint{hep-th/0003059}.

\bibitem[{\citenamefont{Janik}(2001)}]{Janik:2000pp}
\bibinfo{author}{\bibfnamefont{R.~A.} \bibnamefont{Janik}},
  \bibinfo{journal}{Phys.Lett.} \textbf{\bibinfo{volume}{B500}},
  \bibinfo{pages}{118} (\bibinfo{year}{2001}), \eprint{hep-th/0010069}.

\bibitem[{\citenamefont{Polchinski and Strassler}(2002)}]{Polchinski:2001tt}
\bibinfo{author}{\bibfnamefont{J.}~\bibnamefont{Polchinski}} \bibnamefont{and}
  \bibinfo{author}{\bibfnamefont{M.~J.} \bibnamefont{Strassler}},
  \bibinfo{journal}{Phys.Rev.Lett.} \textbf{\bibinfo{volume}{88}},
  \bibinfo{pages}{031601} (\bibinfo{year}{2002}), \eprint{hep-th/0109174}.

\bibitem[{\citenamefont{Polchinski and Strassler}(2003)}]{Polchinski:2002jw}
\bibinfo{author}{\bibfnamefont{J.}~\bibnamefont{Polchinski}} \bibnamefont{and}
  \bibinfo{author}{\bibfnamefont{M.~J.} \bibnamefont{Strassler}},
  \bibinfo{journal}{JHEP} \textbf{\bibinfo{volume}{0305}}, \bibinfo{pages}{012}
  (\bibinfo{year}{2003}), \eprint{hep-th/0209211}.

\bibitem[{\citenamefont{Brower et~al.}(2007)\citenamefont{Brower, Polchinski,
  Strassler, and Tan}}]{Brower:2006ea}
\bibinfo{author}{\bibfnamefont{R.~C.} \bibnamefont{Brower}},
  \bibinfo{author}{\bibfnamefont{J.}~\bibnamefont{Polchinski}},
  \bibinfo{author}{\bibfnamefont{M.~J.} \bibnamefont{Strassler}},
  \bibnamefont{and} \bibinfo{author}{\bibfnamefont{C.-I.} \bibnamefont{Tan}},
  \bibinfo{journal}{JHEP} \textbf{\bibinfo{volume}{0712}}, \bibinfo{pages}{005}
  (\bibinfo{year}{2007}), \eprint{hep-th/0603115}.

\bibitem[{\citenamefont{Brower et~al.}(2009)\citenamefont{Brower, Strassler,
  and Tan}}]{Brower:2007xg}
\bibinfo{author}{\bibfnamefont{R.~C.} \bibnamefont{Brower}},
  \bibinfo{author}{\bibfnamefont{M.~J.} \bibnamefont{Strassler}},
  \bibnamefont{and} \bibinfo{author}{\bibfnamefont{C.-I.} \bibnamefont{Tan}},
  \bibinfo{journal}{JHEP} \textbf{\bibinfo{volume}{0903}}, \bibinfo{pages}{092}
  (\bibinfo{year}{2009}), \eprint{0710.4378}.

\bibitem[{\citenamefont{Brower et~al.}(2010)\citenamefont{Brower, Djuric,
  Sarcevic, and Tan}}]{Brower:2010wf}
\bibinfo{author}{\bibfnamefont{R.~C.} \bibnamefont{Brower}},
  \bibinfo{author}{\bibfnamefont{M.}~\bibnamefont{Djuric}},
  \bibinfo{author}{\bibfnamefont{I.}~\bibnamefont{Sarcevic}}, \bibnamefont{and}
  \bibinfo{author}{\bibfnamefont{C.-I.} \bibnamefont{Tan}},
  \bibinfo{journal}{JHEP} \textbf{\bibinfo{volume}{1011}}, \bibinfo{pages}{051}
  (\bibinfo{year}{2010}), \eprint{1007.2259}.

\bibitem[{\citenamefont{Brower et~al.}(2011)\citenamefont{Brower, Djuric,
  Sarcevic, and Tan}}]{Brower:2011dx}
\bibinfo{author}{\bibfnamefont{R.~C.} \bibnamefont{Brower}},
  \bibinfo{author}{\bibfnamefont{M.}~\bibnamefont{Djuric}},
  \bibinfo{author}{\bibfnamefont{I.}~\bibnamefont{Sarcevic}}, \bibnamefont{and}
  \bibinfo{author}{\bibfnamefont{C.-I.} \bibnamefont{Tan}}
  (\bibinfo{year}{2011}), \eprint{1106.5681}.

\bibitem[{\citenamefont{Hatta et~al.}(2008{\natexlab{a}})\citenamefont{Hatta,
  Iancu, and Mueller}}]{Hatta:2007cs}
\bibinfo{author}{\bibfnamefont{Y.}~\bibnamefont{Hatta}},
  \bibinfo{author}{\bibfnamefont{E.}~\bibnamefont{Iancu}}, \bibnamefont{and}
  \bibinfo{author}{\bibfnamefont{A.}~\bibnamefont{Mueller}},
  \bibinfo{journal}{JHEP} \textbf{\bibinfo{volume}{0801}}, \bibinfo{pages}{063}
  (\bibinfo{year}{2008}{\natexlab{a}}), \eprint{0710.5297}.

\bibitem[{\citenamefont{Hatta et~al.}(2008{\natexlab{b}})\citenamefont{Hatta,
  Iancu, and Mueller}}]{Hatta:2007he}
\bibinfo{author}{\bibfnamefont{Y.}~\bibnamefont{Hatta}},
  \bibinfo{author}{\bibfnamefont{E.}~\bibnamefont{Iancu}}, \bibnamefont{and}
  \bibinfo{author}{\bibfnamefont{A.}~\bibnamefont{Mueller}},
  \bibinfo{journal}{JHEP} \textbf{\bibinfo{volume}{0801}}, \bibinfo{pages}{026}
  (\bibinfo{year}{2008}{\natexlab{b}}), \eprint{0710.2148}.

\bibitem[{\citenamefont{Albacete et~al.}(2008)\citenamefont{Albacete,
  Kovchegov, and Taliotis}}]{Albacete:2008ze}
\bibinfo{author}{\bibfnamefont{J.~L.} \bibnamefont{Albacete}},
  \bibinfo{author}{\bibfnamefont{Y.~V.} \bibnamefont{Kovchegov}},
  \bibnamefont{and} \bibinfo{author}{\bibfnamefont{A.}~\bibnamefont{Taliotis}},
  \bibinfo{journal}{JHEP} \textbf{\bibinfo{volume}{0807}}, \bibinfo{pages}{074}
  (\bibinfo{year}{2008}), \eprint{0806.1484}.

\bibitem[{\citenamefont{Albacete et~al.}(2009)\citenamefont{Albacete,
  Kovchegov, and Taliotis}}]{Albacete:2008vv}
\bibinfo{author}{\bibfnamefont{J.~L.} \bibnamefont{Albacete}},
  \bibinfo{author}{\bibfnamefont{Y.~V.} \bibnamefont{Kovchegov}},
  \bibnamefont{and} \bibinfo{author}{\bibfnamefont{A.}~\bibnamefont{Taliotis}},
  \bibinfo{journal}{AIP Conf.Proc.} \textbf{\bibinfo{volume}{1105}},
  \bibinfo{pages}{356} (\bibinfo{year}{2009}), \eprint{0811.0818}.

\bibitem[{\citenamefont{Basar et~al.}(2012)\citenamefont{Basar, Kharzeev, Yee,
  and Zahed}}]{Basar:2012jb}
\bibinfo{author}{\bibfnamefont{G.}~\bibnamefont{Basar}},
  \bibinfo{author}{\bibfnamefont{D.~E.} \bibnamefont{Kharzeev}},
  \bibinfo{author}{\bibfnamefont{H.-U.} \bibnamefont{Yee}}, \bibnamefont{and}
  \bibinfo{author}{\bibfnamefont{I.}~\bibnamefont{Zahed}},
  \bibinfo{journal}{Phys.Rev.} \textbf{\bibinfo{volume}{D85}},
  \bibinfo{pages}{105005} (\bibinfo{year}{2012}), \eprint{1202.0831}.

\bibitem[{\citenamefont{Stoffers and
  Zahed}(2013{\natexlab{a}})}]{Stoffers:2012zw}
\bibinfo{author}{\bibfnamefont{A.}~\bibnamefont{Stoffers}} \bibnamefont{and}
  \bibinfo{author}{\bibfnamefont{I.}~\bibnamefont{Zahed}},
  \bibinfo{journal}{Phys.Rev.} \textbf{\bibinfo{volume}{D87}},
  \bibinfo{pages}{075023} (\bibinfo{year}{2013}{\natexlab{a}}),
  \eprint{1205.3223}.

\bibitem[{\citenamefont{Stoffers and Zahed}(2012)}]{Stoffers:2012ai}
\bibinfo{author}{\bibfnamefont{A.}~\bibnamefont{Stoffers}} \bibnamefont{and}
  \bibinfo{author}{\bibfnamefont{I.}~\bibnamefont{Zahed}}
  (\bibinfo{year}{2012}), \eprint{1210.3724}.

\bibitem[{\citenamefont{Stoffers and
  Zahed}(2013{\natexlab{b}})}]{Stoffers:2013tla}
\bibinfo{author}{\bibfnamefont{A.}~\bibnamefont{Stoffers}} \bibnamefont{and}
  \bibinfo{author}{\bibfnamefont{I.}~\bibnamefont{Zahed}},
  \bibinfo{journal}{Acta Phys.Polon.Supp.} \textbf{\bibinfo{volume}{6}},
  \bibinfo{pages}{7} (\bibinfo{year}{2013}{\natexlab{b}}).

\bibitem[{\citenamefont{Qian and Zahed}(2012)}]{Zahed:2012sg}
\bibinfo{author}{\bibfnamefont{Y.}~\bibnamefont{Qian}} \bibnamefont{and}
  \bibinfo{author}{\bibfnamefont{I.}~\bibnamefont{Zahed}}
  (\bibinfo{year}{2012}), \eprint{1211.6421}.

\bibitem[{\citenamefont{Shuryak and Zahed}(2014)}]{Shuryak:2013sra}
\bibinfo{author}{\bibfnamefont{E.}~\bibnamefont{Shuryak}} \bibnamefont{and}
  \bibinfo{author}{\bibfnamefont{I.}~\bibnamefont{Zahed}},
  \bibinfo{journal}{Phys.Rev.} \textbf{\bibinfo{volume}{D89}},
  \bibinfo{pages}{094001} (\bibinfo{year}{2014}), \eprint{1311.0836}.

\bibitem[{\citenamefont{Brodsky et~al.}(2013)\citenamefont{Brodsky,
  de~Téramond, and Dosch}}]{Brodsky:2013npa}
\bibinfo{author}{\bibfnamefont{S.~J.} \bibnamefont{Brodsky}},
  \bibinfo{author}{\bibfnamefont{G.~F.} \bibnamefont{de~Téramond}},
  \bibnamefont{and} \bibinfo{author}{\bibfnamefont{H.~G.} \bibnamefont{Dosch}},
  \bibinfo{journal}{Nuovo Cim.} \textbf{\bibinfo{volume}{C036}},
  \bibinfo{pages}{265} (\bibinfo{year}{2013}), \eprint{1302.5399}.

\bibitem[{\citenamefont{Polyakov}(1986)}]{Polyakov:1986cs}
\bibinfo{author}{\bibfnamefont{A.~M.} \bibnamefont{Polyakov}},
  \bibinfo{journal}{Nucl.Phys.} \textbf{\bibinfo{volume}{B268}},
  \bibinfo{pages}{406} (\bibinfo{year}{1986}).

\bibitem[{\citenamefont{Kuti}(2006)}]{Kuti:2005xg}
\bibinfo{author}{\bibfnamefont{J.}~\bibnamefont{Kuti}}, \bibinfo{journal}{PoS}
  \textbf{\bibinfo{volume}{LAT2005}}, \bibinfo{pages}{001}
  (\bibinfo{year}{2006}), \eprint{hep-lat/0511023}.

\bibitem[{\citenamefont{Hidaka and Pisarski}(2009)}]{Hidaka:2009xh}
\bibinfo{author}{\bibfnamefont{Y.}~\bibnamefont{Hidaka}} \bibnamefont{and}
  \bibinfo{author}{\bibfnamefont{R.~D.} \bibnamefont{Pisarski}},
  \bibinfo{journal}{Phys.Rev.} \textbf{\bibinfo{volume}{D80}},
  \bibinfo{pages}{074504} (\bibinfo{year}{2009}), \eprint{0907.4609}.

\bibitem[{\citenamefont{Amaldi and Schubert}(1980)}]{Amaldi:1979kd}
\bibinfo{author}{\bibfnamefont{U.}~\bibnamefont{Amaldi}} \bibnamefont{and}
  \bibinfo{author}{\bibfnamefont{K.~R.} \bibnamefont{Schubert}},
  \bibinfo{journal}{Nucl.Phys.} \textbf{\bibinfo{volume}{B166}},
  \bibinfo{pages}{301} (\bibinfo{year}{1980}).

\bibitem[{\citenamefont{Amaldi et~al.}(1971)\citenamefont{Amaldi, Biancastelli,
  Bosio, Matthiae, Allaby et~al.}}]{Amaldi:1971kt}
\bibinfo{author}{\bibfnamefont{U.}~\bibnamefont{Amaldi}},
  \bibinfo{author}{\bibfnamefont{R.}~\bibnamefont{Biancastelli}},
  \bibinfo{author}{\bibfnamefont{C.}~\bibnamefont{Bosio}},
  \bibinfo{author}{\bibfnamefont{G.}~\bibnamefont{Matthiae}},
  \bibinfo{author}{\bibfnamefont{J.}~\bibnamefont{Allaby}},
  \bibnamefont{et~al.}, \bibinfo{journal}{Phys.Lett.}
  \textbf{\bibinfo{volume}{B36}}, \bibinfo{pages}{504} (\bibinfo{year}{1971}).

\end{thebibliography}
 
Phys.Lett.,B373,210
Phys.Lett.,B429,107

\end{document}